\documentclass[journal]{IEEEtran}

\ifCLASSINFOpdf
\else
   \usepackage[dvips]{graphicx}
\fi
\usepackage{url}

\usepackage{graphicx}
\usepackage{color}
\usepackage{subfig}
\usepackage{booktabs}
\usepackage{enumitem}
\usepackage{wrapfig}
\usepackage{hyperref}
\usepackage{cite}
\usepackage{amsmath,amsfonts,bm}

\def\eqref#1{equation~\ref{#1}}
\def\1{\bm{1}}

\DeclareMathAlphabet{\mathsfit}{\encodingdefault}{\sfdefault}{m}{sl}
\SetMathAlphabet{\mathsfit}{bold}{\encodingdefault}{\sfdefault}{bx}{n}

\newcommand{\R}{\mathbb{R}}

\begin{document}

\title{Self-Supervised Learning of Audio Representations from Permutations with Differentiable Ranking}

\author{Andrew N Carr*\thanks{* BYU CS Department, work done while an intern at Google Brain Paris }\thanks{$\dagger$ Google Brain Paris}, Quentin Berthet$^{\dagger}$, Mathieu Blondel$^{\dagger}$, Olivier Teboul$^{\dagger}$, Neil Zeghidour$^{\dagger}$}

%\markboth{Journal of \LaTeX\ Class Files, Vol. 14, No. 8, August 2015}
\maketitle

\begin{abstract}
Self-supervised pre-training using so-called ``pretext'' tasks has recently shown impressive performance across a wide range of modalities. In this work, we advance self-supervised learning from permutations, by pre-training a model to reorder shuffled parts of the spectrogram of an audio signal, to improve downstream classification performance. 
We make two main contributions.
First, we overcome the main challenges of integrating permutation inversions into an end-to-end training scheme, using recent advances in differentiable ranking. This was heretofore sidestepped by casting the reordering task as classification, fundamentally reducing the space of permutations that can be exploited. Our experiments validate that learning from all possible permutations improves the quality of the pre-trained representations over using a limited, fixed set. Second, we show that inverting permutations is a meaningful pretext task for learning audio representations in an unsupervised fashion. In particular, we improve instrument classification and pitch estimation of musical notes by reordering spectrogram patches in the time-frequency space. 
\end{abstract}


\IEEEpeerreviewmaketitle

\vspace{-1em}
\section{Introduction}

\begin{figure*}[ht]
    \centering
    \includegraphics[scale=0.65]{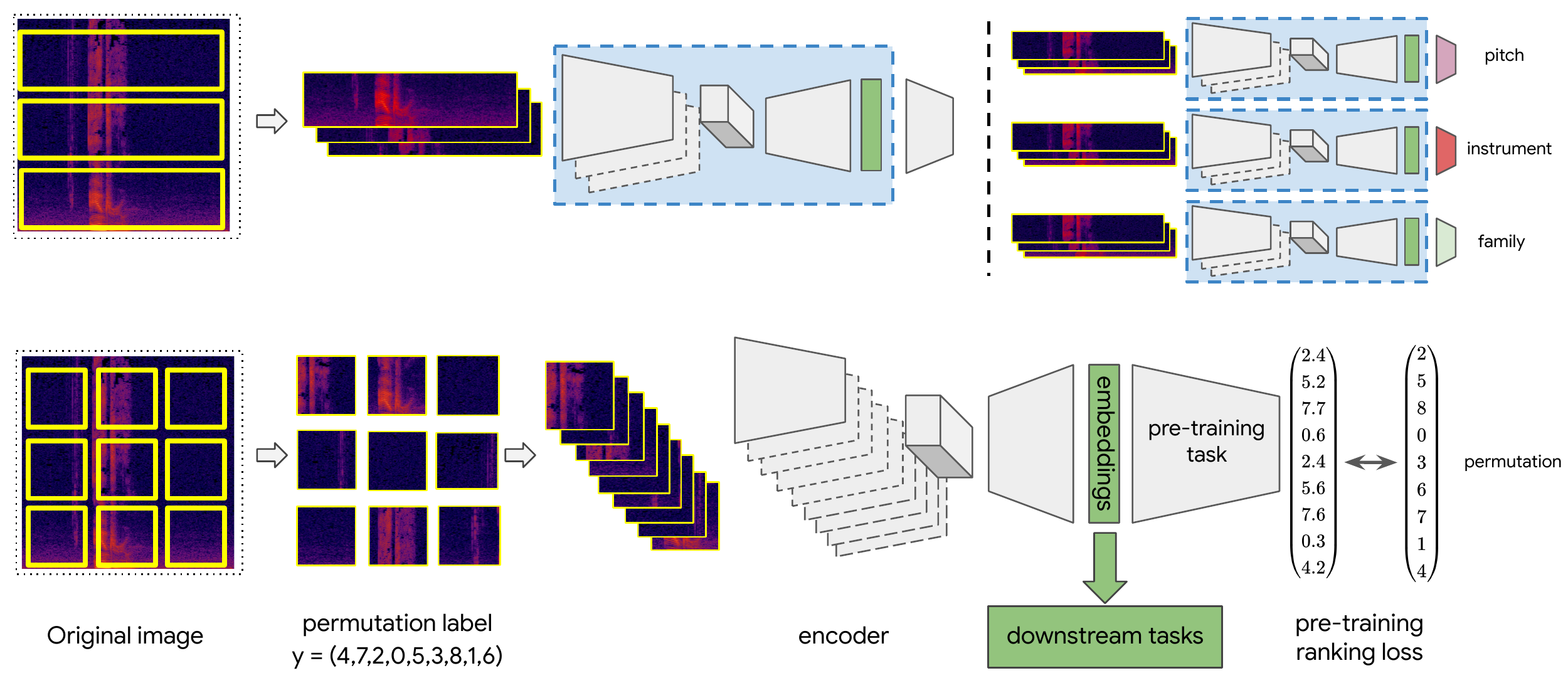}
    \caption{Our framework. {\bf [Bottom]} Patches are generated and permuted on the fly. The network is pre-trained on the associated permutation. Dotted layers indicate weight sharing across input patches. Embeddings used for downstream tasks are extracted by removing the network's last few layers. {\bf [Top]} Downstream training is achieved by freezing the weights of the network up to the embeddings (in blue), and training different shallow classifiers for a variety of tasks.
    Further, permutations as a self-supervised technique can handle a variety of slicing methods with no significant changes to the network architecture.} 
    \label{fig:jigsaw}
\end{figure*}

%Supervised learning has achieved important successes on large annotated datasets \cite{deng2009imagenet, amodei2016deep}. However, most available data, whether images, audio, or videos are unlabelled.
Pre-training representations in an unsupervised way, with subsequent fine-tuning on labelled data, has become the standard to extend the performance of deep architectures to applications where annotations are scarce, such as understanding medical images \cite{rajpurkar2017chexnet}, recognizing speech from under-resourced languages \cite{riviere2020unsupervised, conneau2020unsupervised}, or solving specific language inference tasks \cite{devlin2018bert}. Among unsupervised training schemes, self-supervised learning focuses on designing a proxy training objective, that requires no annotation, such that the representations incidentally learned will generalize well to the task of interest, limiting the amount of labeled data needed for fine-tuning. Such ``pretext'' tasks, a term coined by Doersch et al. \cite{doersch2015unsupervised}, include learning to colorize an artificially gray-scaled image \cite{colorization}, inpainting removed patches \cite{inpainting}, multi-modal data comparisons \cite{ravanelli2020multi}, or recognizing by what angle an original image was rotated \cite{rotation}. Other approaches for self-supervision include classification of original images after data augmentation
\cite{chen2020simple} and clustering \cite{caron2018deep}. Many of these methods, however, such as gray-scale and rotation, cannot be applied to spectrograms for use in audio processing. There has, also been recent work using contrastive learning for audio representation learning \cite{contrastivAudio2020} which requires the raw waveform and log-mel spectrogram. Transformers have also been used \cite{liu2020mockingjay} for unsupervised learning on audio. Our method is beneficial because of the simple architecture, relatively easy change to the loss, and semantically meaningful pre-text task.

In this work, we consider the pretext task of reordering patches of the spectrogram for an audio signal, first proposed for images in \cite{noroozi2016unsupervised}, the analogue of solving a jigsaw puzzle. In this setting, we first split the input into patches and shuffle them by applying a random permutation. We train a neural network to predict which permutation was applied, taking the shuffled patches as inputs. After pre-training with this pretext task, we use the inner representations learned by the neural network as input features to a low-capacity model (see Figure~\ref{fig:jigsaw}) trained for supervised classification (the downstream task). Recent work has used transfer learning on audio to great effect \cite{baevski2020wav2vec}, using self-supervision on the latent vectors, while our work explores the use of a pre-text task directly on the input data. Permutations provide a promising avenue for this type of self-supervised learning, as they are conceptually general enough to be applied across a large range of modalities, unlike colorization \cite{colorization} or rotations \cite{rotation} that are specific to images. The encouraging results of \cite{noroozi2016unsupervised} when transferring learned image features for object detection and image retrieval inspire us to advance this method a step forward. However, including permutations into an end-to-end differentiable pipeline is challenging, as permutations are a discontinuous operation. This issue is circumvented in \cite{noroozi2016unsupervised} casting this problem as classification over a fixed subset of permutations. Given that the number of possible permutations of $n$ patches is $n!$, this approach cannot scale to the full set of permutations, even when $n$ is moderately small. Alternatively, \cite{santa2017deeppermnet} use Sinkhorn normalization to produce doubly stochastic matrices as approximations of the true permutation matrix. While this method can be integrated into an end-to-end pipeline, each forward pass through the model relies on an iterative Sinkhorn algorithm, of which each iteration has a cost $O(n^2)$ and which requires choosing a stopping criterion.

In this work, we leverage recent advances in differentiable ranking \cite{berthet2020learning, blondel2020fast} to integrate permutations into end-to-end neural training. This allows us to solve the permutation inversion task for the entire set of permutations, removing a bottleneck that was heretofore sidestepped in manners that limits downstream performance. Moreover, we demonstrate for the first time the effectiveness of permutations as a pretext task on audio with minimal modality-specific adjustments. In particular, we improve instrument classification and pitch estimation of musical notes by learning to reorder spectrogram frames, over the time and frequency axes.

The rest of the paper is organized as follows. In Section~\ref{sec:methodo} we present the problem formulation and methods. In Section~\ref{sec:experiments} we demonstrate the effectiveness of our system on instrument classification and pitch estimation of musical notes.

\section{Methods}
\label{sec:methodo}

\vspace{-0.25em}
\subsection{General methodology}

In this section, we present a self-supervised pretext task that predicts the permutation applied to patches of an input. We do so in a manner that allows to use all possible permutations as targets during training. This pretext task is performed for {\em pre-training} and the internal representation learned by the pretext neural network can be transferred and used on secondary {\em downstream} tasks -- see Figure \ref{fig:jigsaw}.  

%We address the problem of self-supervised learning of representations, using a large unlabelled dataset. 

During pre-training, for each audio sequence, we split its spectrogram into $n$ {\em patches}. These patches can either be vertical (stacks of time frames), horizontal (stacks of frequency bands), or on both axes (stacks of time-frequency bins). %Their exact structure can depend on the desired task. Accordingly, the dimensions in $d$ can represent height, width, channels, etc. 
We then permute these patches randomly, and stack them in a tensor $X_i$ of dimension $n \times d$ (see Figure~\ref{fig:jigsaw}), which is paired to the applied permutation as a label $y_i$ of size $n$ (see Section~\ref{sec:theory} for details on permutation encoding).

%+In our work, the upstream task (which uses permutations) is a pretext task defined over a set of unlabeled data $X \in \mathbb{R}^d$. Labels are drawn uniformly from the permutahedron, which is a simplex whose vertices are permutations. The unlabeled data is then paired with a random permutation (which can change between epochs). This pairing process involves slicing the image into set of patches corresponding to elements of the chosen permutation. 

We pre-train the weights $w$ of a neural network to invert this permutation, using a differentiable ranking operator $y^*_\varepsilon$. This operator, and other details of pre-training are described in Section~\ref{sec:theory}; the network and data-processing are described in Section~\ref{sec:applications}. After pre-training, the network weights are used to generate embeddings at an intermediate layer. These representations can be used in a {\em downstream} task, as input to a low-capacity classifier (see Figure \ref{fig:jigsaw}).

%After pre-training on the initial dataset, the network weights can be used to generate representations, generating embeddings at an intermediate layer. These representations can be used in a {\em downstream} task to train a new network, with its own weights, minimizing a loss (typically classification or regression) between its output and the downstream task labels (see Figures ~\ref{fig:jigsaw} and~\ref{fig:full_process}). 

We mostly evaluate our methods by improvements in downstream classification. However, the reordering task can be of interest in itself, as in learning-to-rank problems \cite{liu2011learning}, and we also report generalization performance in this task.

\subsection{Differentiable ranking methodology}
\label{sec:theory}

Our methodology for representation learning relies on the ability to incorporate ordering or ranking operations in an end-to-end differentiable pipeline. This is achieved by using a convenient encoding of the permutations of $n$ objects in $\R^n$, and differentiable operators that approximate ranking.

For each permutation, the label $y_i$ is a vector of ranks, or relative order, of the elements (e.g. $y=(0, 1, 2, 3)$ for the identity permutation). For the model prediction, the last two layers consist of: a vector of score values $\theta_w(X) \in \R^n$, and network outputs $f_w(X) = y^*_\varepsilon( \theta_w(X)) \in \R^n$, using differentiable ranking operators $y^*_\varepsilon$. 

These operations map any vector of $n$ values to a point in the convex hull of permutation encodings in dimension $n$ (e.g. $(0.1,0.9,2.2,2.8)$ over 4 elements), akin to a softmax operator in a classification setting. We consider here two differentiable ranking operations, either using stochastic perturbations \cite{berthet2020learning} or regularization \cite{blondel2020fast}. In any case, embedding the permutations in this manner (rather than e.g. permutation matrices or $n!$ classes) puts more emphasis on the relative position of the elements, and enables us to penalize less smaller index differences.

These tools ensure that our model is end-to-end differentiable, and enables to use all permutations in training. This is unlike the models of  \cite{noroozi2016unsupervised} and \cite{lee2017unsupervised}, where reordering is reduced to classification, assigning a set of $L$ permutations to one-hot vectors in $\R^L$. This approach is obviously limited: representing all the permutations requires in principle $n!$ classes, which is quickly not manageable, even for small values of $n$.

Pre-training the network parameters $w$ requires a loss function between $y_\varepsilon^*(\theta)$ and $y$. For the version of $y_\varepsilon^*(\theta)$ based on stochastic perturbations, we use the associated {\em Fenchel--Young loss} (``Perturbed F-Y'' in empirical results) \cite{blondel2020learning}, that acts directly on $\theta = \theta_w(X)$ written here as $L_{\sf FY}(\theta;y)$. Its gradients, given by $\nabla_\theta L_{\sf FY}(\theta;y) = y_\varepsilon^*(\theta) - y$, are easy to compute. For the regularized version of $y_\varepsilon^*(\theta)$ \cite{blondel2020fast}, we use $\|y_\varepsilon^*(\theta) - y\|^2/2$.
(``Fast Soft Ranking'' in  empirical results).

We opt for these two losses for their good theoretical properties and $O(n \log n)$ complexity. Other choices 
\cite{santa2017deeppermnet, mena2018learning,cuturi2019differentiable,vlastelica2019differentiation,rolinek2020optimizing,grover2019stochastic} are also possible, potentially with higher computational cost, or regions with zero gradient.% In particular, our methods are significantly more efficient than the algorithm of \cite{}, which uses an iterative algorithm of which each iteration has a $O(n^2)$ cost.

\vspace{-1em}

\subsection{Implementation and architecture}
\label{sec:applications}

\paragraph{Data-processing}
When constructing the self-supervised task, we slice inputs in patches. This slicing is controlled by two variables $n_x$ and $n_y$, determining respectively the number of columns and rows used. In \cite{noroozi2016unsupervised}, $9$ square patches are used for images. On spectrograms, this choice is conceptually richer as using $n_x = 1$ means slicing frequency bands, while using $n_y = 1$ means slicing along the time axis. Using both $n_x \neq 1$ and $n_y \neq 1$ allows slicing along both axes, see Figure \ref{fig:jigsaw} for an illustration.
%Therefore, if we wish to slice along the frequency domain in a spectrogram, we let $nx = 1$ and $ny = 3$ as shown in (fig \ref{fig:jigsaw}).

\paragraph{Pre-training task}

For the reordering pretext task, we use a Context Free Network (CFN) from  \cite{noroozi2016unsupervised}. This network uses an AlexNet \cite{krizhevsky2012imagenet} backbone which processes each patch individually while sharing weights across all patches as shown in Figure \ref{fig:jigsaw}. By processing each patch independently, but with shared weights, the network cannot rely on global structure. After the shared backbone, the patches are passed together through two fully connected layers. %The final layer of our CFN variant is determined by the number of patches in the permutation (e.g. $n_x \times n_y$ for images).
The output layer represents the predicted ranks of the input permutation.

% One potential reason why the MSE also performs better than the original fixed permutation classification work is due to the geometry of the output space. In other words, because the MSE measures distance directly between vertices of the permutahedron it can differentiate between two vertices that only differ by a single entry. Whereas the fixed permutation classification scheme uses one-hot vectors, which cannot differentiate between two similar vertices in the same way. 

\paragraph{Downstream task}

In the downstream tasks we use 3-layer multi-layer perceptrons (MLP) trained on embeddings extracted at the first aggregate fully connected layer of the pretext network (whose weights are frozen during this part). The MLP's output layer is task-dependent. For a regression downstream task, the output of the MLP is a single scalar and the downstream model is trained to minimize mean-squared error. For classification, the output of the MLP is a softmax over the class logits, and we train the downstream model by minimizing the cross-entropy loss.

\begin{figure*}
    \centering
    \hbox{
    \hspace{-1.3em}
    \includegraphics[width=\textwidth]{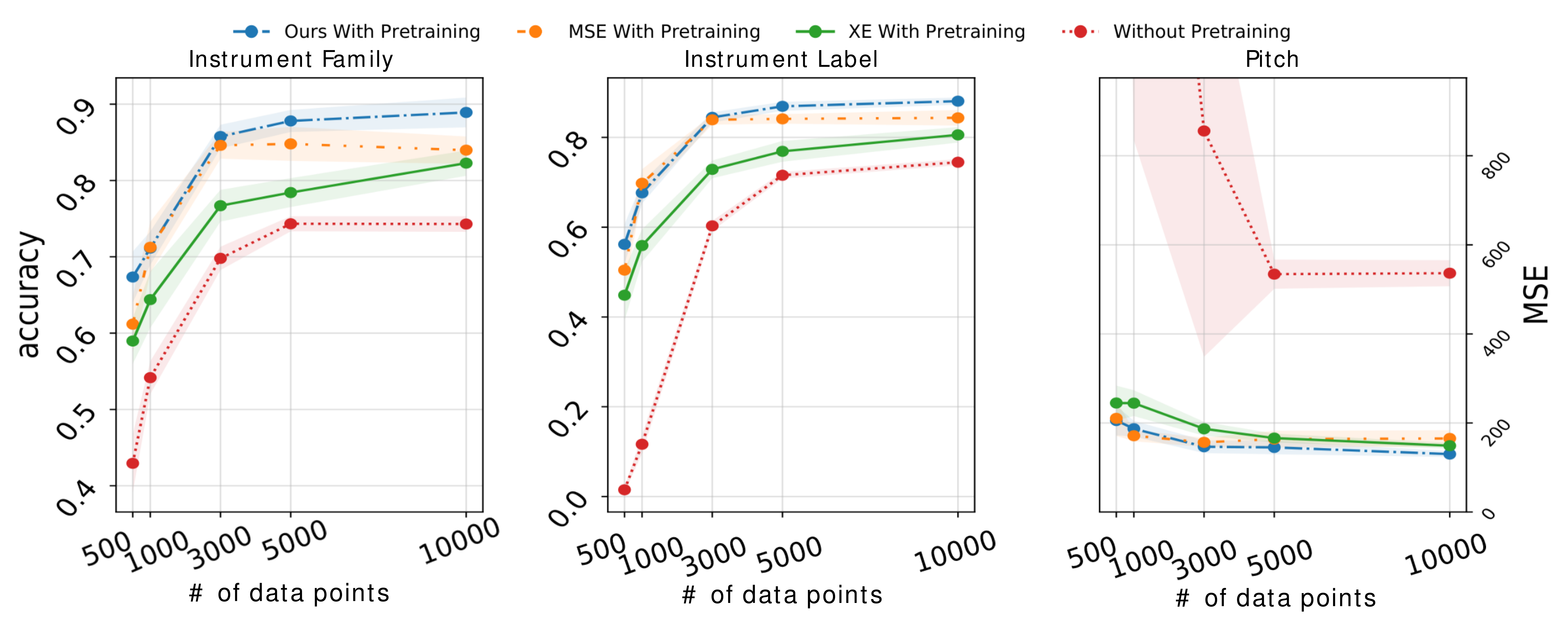}
    }
    \vspace{-1em}
    \caption{Performance of our permutation-based pretraining over 3 audio tasks when varying the number of data points in the downstream task. Higher is better for Instrument Family and Instrument Label while lower is better in Pitch prediction.}
    \label{fig:dataset_comparison}
\end{figure*}

\section{Experiments}
\label{sec:experiments}

We demonstrate the effectiveness of permutation-based pre-training as measured by the test accuracy in instrument classification and pitch estimation tasks.
All experiments are carried out in Tensorflow \cite{tensorflow} and run on a single P100 GPU.
We also report the performance on the pre-training task using partial ranks, the proportion of patches ranked in the correct position.
%We will open-source our codebase for reproducibility and reusability.

%The experiments run and presented are designed to give evidence to our main hypothesis that extending self-supervised methods to use all available permutations improves the quality of learned representations as measured by generalization in downstream task performance (i.e on a downstream test set). All experiments were written in TensorFlow and run on a single P100 GPU.

\vspace{-1em}
\subsection{Experimental setup.}
The NSynth dataset \cite{nsynth2017} offers about 300,000 audio samples of musical notes, each with a unique pitch, timbre, and envelope recorded from 1,006 different instruments. The recordings, sampled at 16kHz, are 4 seconds long and can be used for 3 downstream classification tasks: predicting the instrument itself (1,006 classes) the instrument family (11 classes) and predicting the pitch of the note (128 values). We formulate pitch estimation as a regression task and report the mean squared error (MSE). The input representation is a log-compressed spectrogram, computed over 25ms with a 10ms stride and 513 bins. The 2D structure of the spectrogram allows us to use a 2D convolutional neural network, as is done with images. We train our CFN with an AlexNet backbone on the pre-training task of predicting applied permutations for 1000 epochs, over mini batches of size 32 and with an Adam optimizer \cite{kingma2014adam} with a learning rate of $10^{-6}$. We then evaluate the downstream generalization performance over the 3 NSynth tasks, by replacing the last layers of the network by a task specific 3-layer MLP and replacing the random permutation by the identity.

% To understand the quality of the produced embeddings, we vary the number of examples used to train train the downstream task and report the results for different data regimes.

We compare the different variants of our method (number and nature of the patches) with 2 baseline alternatives: i) training the downstream head on an untrained encoder (Random Embedding) and  ii) solving the same pre-training task but using instead a finite set of 100 permutations as proposed by \cite{noroozi2016unsupervised} (Fixed Permutation). We also compare different losses to train the permutation pretext task: a) cross entropy (XE) when learning over 100 permutations, b) MSE loss (MSE), c) soft ranking via perturbations (Perturbed F-Y) and d) soft ranking (Fast Soft Ranking).

\vspace{-1em}
\subsection{Empirical results.}

First, we compare the different methods across several downstream data regimes and report the results in Figure~\ref{fig:dataset_comparison}. Here, all pre-training models slice the spectrogram over the frequency axis as it gives the best performance (see \ref{subsection:ablation} for an ablation study). Additionally, the experiments in this figure use 10 patches for all methods and 1000 permutations for the XE method. We observe that in the low data regime our method strongly outperforms an end-to-end fully supervised model. Moreover, this advantage is maintained when increasing the number of training examples in the downstream tasks. We also observe that pretraining is particularly impactful for pitch estimation which aligns with results of \cite{gfeller2020spice}.
\begin{figure*}
    \centering
    \vspace{-1em}
    \hbox{
    \hspace{-1.3em}
    \includegraphics[width=\textwidth]{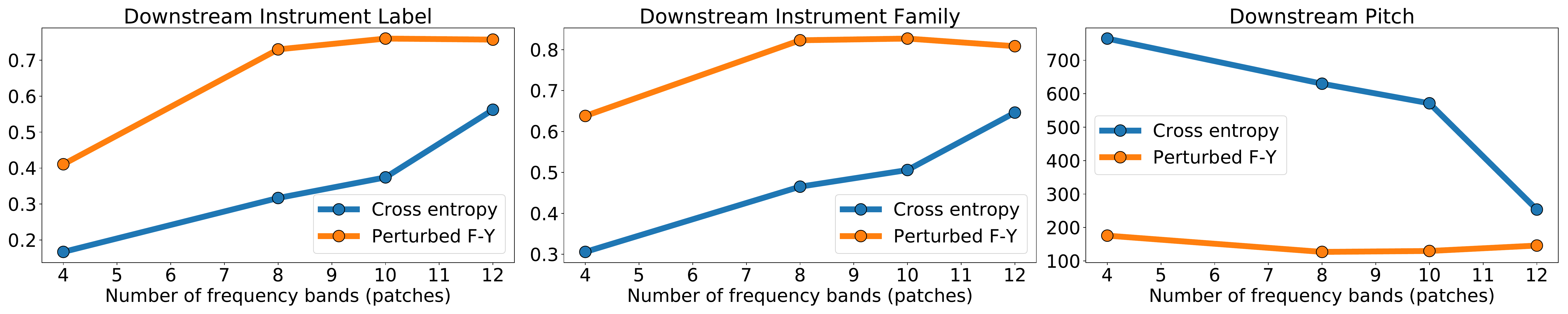}
    }
    \caption{Performance on the downstream tasks, as a function of the number of frequency patches used for pre-training.}
    \label{fig:performance_num_patches}
\end{figure*}

\begin{table}
\centering
\setlength\tabcolsep{2pt}
\resizebox{\columnwidth}{!}{%
\begin{tabular}{lccccc}
\toprule
 Task  & Random & Fixed & Fast & Perturbed & MSE \\
 &  Embedding & Permutation & Soft Ranking & F-Y & \\
\hline
Instr. Family (ACC) & $0.46 $ {\tiny $  \pm   0.01$}  & $0.75 $ {\tiny $  \pm   0.02$}  &    $0.84 $ {\tiny $ \pm   0.02$} &    $\bm{0.85} $ {\tiny $ \pm   0.03$} &    $0.79 $ {\tiny $\pm   0.02$} \\
Instr. Label (ACC) & $0.35  $ {\tiny $ \pm   0.04$} & $0.70  $ {\tiny $  \pm   0.03$} &    $0.72 $ {\tiny $ \pm   0.03$} &    $\bm{0.76} $ {\tiny $ \pm   0.04$} &    $0.71 $ {\tiny $\pm   0.03$} \\
Pitch (MSE)   & $258.76 $ {\tiny $ \pm  3$} & $144.48 $ {\tiny $ \pm  18$} &    $133.85 $ {\tiny $ \pm 19$} &    $\bm{124.6} $ {\tiny $ \pm 14$} &   $156.08 $ {\tiny $ \pm 21$} \\
\midrule
Partial Ranks Accuracy   & - & - & $0.57 $ {\tiny $ \pm  0.0$} & $\bm{0.58}$ {\tiny $ \pm  0.0$} &    $0.51$ {\tiny $ \pm 0.0$}\\
%hline                                            
\bottomrule
\end{tabular}
}
\caption{\label{tab:audio_downstream_1000}Performance on three downstream tasks with 1000 downstream data points taken from NSynth. ACC stands for accuracy, and MSE for mean squared error.}
\vspace{-2em}
\end{table}

% (see \ref{tab:audio_downstream_500} and \ref{tab:audio_downstream_5000}, in the Appendix \ref{sec:appresults} for 500 and 5000).

We report in Table~\ref{tab:audio_downstream_1000} the results for the the three downstream tasks, using 1000 training examples. Those experiments are run using 10 frequency bands, which corresponds to $10!$ permutations. In this context, casting permutation inversion as a classification problem is limited to exploiting only a small proportion of possible permutations, as it would otherwise amount to 3.6M classes. On the other hand, our method scales very well in this setting.
We first observe that random embeddings perform poorly but do represent a good baseline to be compared to, as the difference in performance with other methods illustrates what is gained from pre-training. Second, the baseline is significantly outperformed when using a fixed set of permutation and a classification loss, as in \cite{noroozi2016unsupervised}. We then observe that even with a mean squared error loss, the performance on the downstream task is comparable or better than the fixed permutation method and we show that using a ranking loss further increases the performance. Furthermore, Fig.\ref{fig:performance_num_patches} shows the effect of the number of frequency bands on the downstream performance. As the number of permutations grows, the overall performance over the downstream task increases, providing better representations, up to 9-10 patches. These results tend to confirm that i) permutation is an interesting pretext task, ii) considering all possible permutation helps building better representations and iii) the use of a ranking loss is the right choice of loss for such tasks. We then report in the last row of Table \ref{tab:audio_downstream_1000} performance in the pretext task. Good performance on the downstream task is often connected to good performance on the pretext task. Here, we measure performance by the ability of the CFN network to reorder the shuffled inputs, reporting the proportion of items ranked in the correct position.

% We observe that the performance of the fully differentiable ranking methods is higher than a direct MSE comparison when using 10 patches. We also observed that with lower number of patches, the upstream task can become too easy to lead to good feature learning: there is a risk of overfitting with embeddings of lower quality as measured by downstream task performance. Therefore, there is a balance between pretext task difficulty and downstream performance. We found for the Nsynth task that 10 patches ($10!$ potential permutations, more than 3.6 million) along the frequency domain was the best performing configuration when taking this balance into account.

%\ndqb{shorten following} When we train on 10 patches, we observe that the performance of the fully differentiable ranking methods is higher than a direct MSE comparison. Interestingly, however, we noticed that if the number of patches is decreased too far then the embeddings learned by the CFN in the . 

\begin{table}
\begin{center}
\setlength\tabcolsep{2pt}
\begin{tabular}{llll}
\toprule
Task &            Frequency &                    Time &     Time-Frequency\\
\hline                                           
Instr. Family (ACC) & $    \bm{0.82}$ {\tiny $  \pm 0.003$} & $0.79 $ {\tiny $  \pm   0.004 $} & $0.73  $ {\tiny $ \pm   0.003$} \\
Instr. Label (ACC) & $    \bm{0.75} $ {\tiny $ \pm 0.01$} & $0.68 $ {\tiny $  \pm   0.002$} & $0.45 $ {\tiny $  \pm   0.01$} \\
Pitch (MSE)   & $ \bm{136.68} $ {\tiny $ \pm 12.63$}           & $206.11 $ {\tiny $  \pm  7.32$} & $\bm{137.81} $ {\tiny $  \pm  13.02$} \\
\bottomrule
\end{tabular}
\caption{\label{tab:vert_horz_sq}Slicing biases on Nsynth downstream tasks.}
\end{center}
\vspace{-2.5em}
\end{table}

\vspace{-1em}
\subsection{Time-frequency structure and permutations}
\label{subsection:ablation}
Unlike images, the horizontal and vertical dimensions of a spectrogram are semantically different, respectively representing time and frequency. While \cite{noroozi2016unsupervised} only exploited square patches, experimenting with audio allows exploring permutations over frequency bands (horizontal patches), time frames (vertical patches) or square time-frequency patches, and comparing the resulting downstream performance.
Table \ref{tab:vert_horz_sq} reports a comparison between these three settings. Overall, shuffling along the frequency axis is the best pre-training strategy. These results illustrate a specificity of the dataset: our inputs are single notes, many of them having an harmonic structure. In this context, learning the relation between frequency bands is meaningful both to recognize which instrument is playing, as well as which note (pitch) is being played. This also explains the poor performance of slicing along the time axis. Pitch is a time-independent characteristic, so the time structure is not relevant for this task. Moreover, musical notes have an easily identifiable time structure (fast attack and slow decay), which may make the task of reordering time frames trivial. We hypothesize that signals with a richer, non-stationary time structure, such as speech, would benefit more from shuffling time frames.

\vspace{-1em}
\section{Conclusion}
\label{sec:conclusion}

We present a general pre-training method that uses permutations to learn high-quality representations from spectrograms, and improves the downstream performance of audio classification and regression tasks on musical notes. We demonstrate that our method outperforms previous permutation learning schemes by incorporating fully differentiable ranking as a pretext loss, enabling us to take advantage of all $n!$ permutations, instead of a small fixed set. In particular, we show significant improvements in low data regimes.

% \section*{Acknowledgment}

\bibliographystyle{IEEEtran}
\bibliography{spl}

\begin{thebibliography}{10}
\providecommand{\url}[1]{#1}
\csname url@samestyle\endcsname
\providecommand{\newblock}{\relax}
\providecommand{\bibinfo}[2]{#2}
\providecommand{\BIBentrySTDinterwordspacing}{\spaceskip=0pt\relax}
\providecommand{\BIBentryALTinterwordstretchfactor}{4}
\providecommand{\BIBentryALTinterwordspacing}{\spaceskip=\fontdimen2\font plus
\BIBentryALTinterwordstretchfactor\fontdimen3\font minus
  \fontdimen4\font\relax}
\providecommand{\BIBforeignlanguage}[2]{{%
\expandafter\ifx\csname l@#1\endcsname\relax
\typeout{** WARNING: IEEEtran.bst: No hyphenation pattern has been}%
\typeout{** loaded for the language `#1'. Using the pattern for}%
\typeout{** the default language instead.}%
\else
\language=\csname l@#1\endcsname
\fi
#2}}
\providecommand{\BIBdecl}{\relax}
\BIBdecl

\bibitem{rajpurkar2017chexnet}
P.~Rajpurkar, J.~Irvin, K.~Zhu, B.~Yang, H.~Mehta, T.~Duan, D.~Ding, A.~Bagul,
  C.~Langlotz, K.~Shpanskaya \emph{et~al.}, ``Chexnet: Radiologist-level
  pneumonia detection on chest x-rays with deep learning,'' \emph{arXiv
  preprint arXiv:1711.05225}, 2017.

\bibitem{riviere2020unsupervised}
M.~Rivi{\`e}re, A.~Joulin, P.-E. Mazar{\'e}, and E.~Dupoux, ``Unsupervised
  pretraining transfers well across languages,'' in \emph{ICASSP 2020-2020 IEEE
  International Conference on Acoustics, Speech and Signal Processing
  (ICASSP)}.\hskip 1em plus 0.5em minus 0.4em\relax IEEE, 2020, pp. 7414--7418.

\bibitem{conneau2020unsupervised}
A.~Conneau, A.~Baevski, R.~Collobert, A.~Mohamed, and M.~Auli, ``Unsupervised
  cross-lingual representation learning for speech recognition,'' \emph{arXiv
  preprint arXiv:2006.13979}, 2020.

\bibitem{devlin2018bert}
J.~Devlin, M.-W. Chang, K.~Lee, and K.~Toutanova, ``Bert: Pre-training of deep
  bidirectional transformers for language understanding,'' \emph{arXiv preprint
  arXiv:1810.04805}, 2018.

\bibitem{doersch2015unsupervised}
C.~Doersch, A.~Gupta, and A.~A. Efros, ``Unsupervised visual representation
  learning by context prediction,'' in \emph{Proceedings of the IEEE
  international conference on computer vision}, 2015, pp. 1422--1430.

\bibitem{colorization}
G.~Larsson, M.~Maire, and G.~Shakhnarovich, ``Colorization as a proxy task for
  visual understanding,'' in \emph{Proceedings of the IEEE Conference on
  Computer Vision and Pattern Recognition}, 2017, pp. 6874--6883.

\bibitem{inpainting}
D.~Pathak, P.~Krahenbuhl, J.~Donahue, T.~Darrell, and A.~A. Efros, ``Context
  encoders: Feature learning by inpainting,'' in \emph{Proceedings of the IEEE
  conference on computer vision and pattern recognition}, 2016, pp. 2536--2544.

\bibitem{ravanelli2020multi}
M.~Ravanelli, J.~Zhong, S.~Pascual, P.~Swietojanski, J.~Monteiro, J.~Trmal, and
  Y.~Bengio, ``Multi-task self-supervised learning for robust speech
  recognition,'' in \emph{ICASSP 2020-2020 IEEE International Conference on
  Acoustics, Speech and Signal Processing (ICASSP)}.\hskip 1em plus 0.5em minus
  0.4em\relax IEEE, 2020, pp. 6989--6993.

\bibitem{rotation}
S.~Gidaris, P.~Singh, and N.~Komodakis, ``Unsupervised representation learning
  by predicting image rotations,'' \emph{arXiv preprint arXiv:1803.07728},
  2018.

\bibitem{chen2020simple}
T.~Chen, S.~Kornblith, M.~Norouzi, and G.~Hinton, ``A simple framework for
  contrastive learning of visual representations,'' \emph{arXiv preprint
  arXiv:2002.05709}, 2020.

\bibitem{caron2018deep}
M.~Caron, P.~Bojanowski, A.~Joulin, and M.~Douze, ``Deep clustering for
  unsupervised learning of visual features,'' in \emph{Proceedings of the
  European Conference on Computer Vision (ECCV)}, 2018, pp. 132--149.

\bibitem{contrastivAudio2020}
L.~Wang and A.~van~den Oord, ``Multi-format contrastive learning of audio
  representations,'' \emph{NeurIPS Workshops (Self-Supervised Learning for
  Speech and Audio Processing)}, 2020.

\bibitem{liu2020mockingjay}
A.~T. Liu, S.-w. Yang, P.-H. Chi, P.-c. Hsu, and H.-y. Lee, ``Mockingjay:
  Unsupervised speech representation learning with deep bidirectional
  transformer encoders,'' in \emph{ICASSP 2020-2020 IEEE International
  Conference on Acoustics, Speech and Signal Processing (ICASSP)}.\hskip 1em
  plus 0.5em minus 0.4em\relax IEEE, 2020, pp. 6419--6423.

\bibitem{noroozi2016unsupervised}
M.~Noroozi and P.~Favaro, ``Unsupervised learning of visual representations by
  solving jigsaw puzzles,'' in \emph{European Conference on Computer
  Vision}.\hskip 1em plus 0.5em minus 0.4em\relax Springer, 2016, pp. 69--84.

\bibitem{baevski2020wav2vec}
A.~Baevski, Y.~Zhou, A.~Mohamed, and M.~Auli, ``wav2vec 2.0: A framework for
  self-supervised learning of speech representations,'' \emph{Advances in
  Neural Information Processing Systems}, vol.~33, 2020.

\bibitem{santa2017deeppermnet}
R.~Santa~Cruz, B.~Fernando, A.~Cherian, and S.~Gould, ``Deeppermnet: Visual
  permutation learning,'' in \emph{Proceedings of the IEEE Conference on
  Computer Vision and Pattern Recognition}, 2017, pp. 3949--3957.

\bibitem{berthet2020learning}
Q.~Berthet, M.~Blondel, O.~Teboul, M.~Cuturi, J.-P. Vert, and F.~Bach,
  ``Learning with differentiable perturbed optimizers,'' \emph{arXiv preprint
  arXiv:2002.08676}, 2020.

\bibitem{blondel2020fast}
M.~Blondel, O.~Teboul, Q.~Berthet, and J.~Djolonga, ``Fast differentiable
  sorting and ranking,'' \emph{arXiv preprint arXiv:2002.08871}, 2020.

\bibitem{liu2011learning}
T.-Y. Liu, \emph{Learning to rank for information retrieval}.\hskip 1em plus
  0.5em minus 0.4em\relax Springer Science \& Business Media, 2011.

\bibitem{lee2017unsupervised}
H.-Y. Lee, J.-B. Huang, M.~Singh, and M.-H. Yang, ``Unsupervised representation
  learning by sorting sequences,'' in \emph{Proceedings of the IEEE
  International Conference on Computer Vision}, 2017, pp. 667--676.

\bibitem{blondel2020learning}
M.~Blondel, A.~F. Martins, and V.~Niculae, ``Learning with fenchel-young
  losses.'' \emph{Journal of Machine Learning Research}, vol.~21, no.~35, pp.
  1--69, 2020.

\bibitem{mena2018learning}
G.~Mena, D.~Belanger, S.~Linderman, and J.~Snoek, ``Learning latent
  permutations with gumbel-sinkhorn networks,'' \emph{arXiv preprint
  arXiv:1802.08665}, 2018.

\bibitem{cuturi2019differentiable}
M.~Cuturi, O.~Teboul, and J.-P. Vert, ``Differentiable ranking and sorting
  using optimal transport,'' in \emph{Advances in Neural Information Processing
  Systems}, 2019, pp. 6861--6871.

\bibitem{vlastelica2019differentiation}
M.~Vlastelica, A.~Paulus, V.~Musil, G.~Martius, and M.~Rol{\'\i}nek,
  ``Differentiation of blackbox combinatorial solvers,'' \emph{arXiv preprint
  arXiv:1912.02175}, 2019.

\bibitem{rolinek2020optimizing}
M.~Rol{\'\i}nek, V.~Musil, A.~Paulus, M.~Vlastelica, C.~Michaelis, and
  G.~Martius, ``Optimizing rank-based metrics with blackbox differentiation,''
  in \emph{Proceedings of the IEEE/CVF Conference on Computer Vision and
  Pattern Recognition}, 2020, pp. 7620--7630.

\bibitem{grover2019stochastic}
A.~Grover, E.~Wang, A.~Zweig, and S.~Ermon, ``Stochastic optimization of
  sorting networks via continuous relaxations,'' \emph{arXiv preprint
  arXiv:1903.08850}, 2019.

\bibitem{krizhevsky2012imagenet}
A.~Krizhevsky, I.~Sutskever, and G.~E. Hinton, ``Imagenet classification with
  deep convolutional neural networks,'' in \emph{Advances in neural information
  processing systems}, 2012, pp. 1097--1105.

\bibitem{tensorflow}
M.~Abadi, P.~Barham, J.~Chen, Z.~Chen, A.~Davis, J.~Dean, M.~Devin,
  S.~Ghemawat, G.~Irving, M.~Isard \emph{et~al.}, ``Tensorflow: A system for
  large-scale machine learning,'' in \emph{12th $\{$USENIX$\}$ symposium on
  operating systems design and implementation ($\{$OSDI$\}$ 16)}, 2016, pp.
  265--283.

\bibitem{nsynth2017}
J.~Engel, C.~Resnick, A.~Roberts, S.~Dieleman, D.~Eck, K.~Simonyan, and
  M.~Norouzi, ``Neural audio synthesis of musical notes with wavenet
  autoencoders,'' 2017.

\bibitem{kingma2014adam}
D.~P. Kingma and J.~Ba, ``Adam: A method for stochastic optimization,''
  \emph{arXiv preprint arXiv:1412.6980}, 2014.

\bibitem{gfeller2020spice}
B.~Gfeller, C.~Frank, D.~Roblek, M.~Sharifi, M.~Tagliasacchi, and
  M.~Velimirovi{\'c}, ``Spice: Self-supervised pitch estimation,''
  \emph{IEEE/ACM Transactions on Audio, Speech, and Language Processing},
  vol.~28, pp. 1118--1128, 2020.

\end{thebibliography}
%\printbibliography
% \bibliographystyle{iclr2021_conference}

\end{document}